\documentclass[
 reprint,twocolumn,
 amsmath,amssymb,
 aps,
]{revtex4-1}

\usepackage{epsfig,bm,epsf,graphics}
\usepackage{graphicx}
\usepackage{dcolumn}
\usepackage{bm}

\begin{document}
\preprint{APS/123-QED}

\title{Stability and Phase Transition of Skyrmion Crystals Generated by Dzyaloshinskii-Moriya Interaction}

\author{Sahbi El Hog\footnote{sahbi.el-hog@u-cergy.fr}, Aur\'elien Bailly-Reyre \footnote{aurelien.bailly-reyre@u-cergy.fr} and H. T. Diep\footnote{diep@u-cergy.fr, corresponding author}}
\affiliation{%
Laboratoire de Physique Th\'eorique et Mod\'elisation,
Universit\'e de Cergy-Pontoise, CNRS, UMR 8089\\
2, Avenue Adolphe Chauvin, 95302 Cergy-Pontoise Cedex, France.\\
 }%

\date{\today}

\begin{abstract}
We generate a crystal of skyrmions in two dimensions using a Heisenberg Hamiltonian including the ferromagnetic interaction $J$, the Dzyaloshinskii-Moriya interaction $D$, and an applied magnetic field $H$.  The ground state (GS) is determined by minimizing the interaction energy. We show that the GS is a skyrmion crystal in a region of $(D,H)$. The stability of this skyrmion crystalline phase at finite temperatures is shown by a study of the time-dependence of the order parameter using Monte Carlo simulations. We observe that the relaxation is very slow and follows a stretched exponential law.  The skyrmion crystal phase is shown to undergo a transition to the paramagnetic state at a finite temperature.
\vspace{0.5cm}
\begin{description}
\item[PACS numbers: 75.25.-j ; 75.30.Ds ; 75.70.-i ]
\end{description}
\end{abstract}

\pacs{Valid PACS appear here}
\maketitle


\section{Introduction}

Skyrmions have been extensively investigated in condensed matter physics \cite{Fert2013} since its theoretical formulation by T. H. R. Skyrme \cite{Skyrme} in the context of nuclear matter.

There are several mechanisms and interactions leading to the appearance of skyrmions in various kinds of matter. The most popular one is certainly the Dzyaloshinskii-Moriya (DM) interaction which was initially proposed to explain the weak ferromagnetism observed in antiferromagnetic Mn compounds.   The phenomenological Landau-Ginzburg model introduced by I. Dzyaloshinskii \cite{Dzyaloshinskii} was microscopically derived by T. Moriya \cite{Moriya}.
This demonstration shows that the DM interaction comes from the second-order perturbation of the exchange interaction between two spins which is not zero only under some geometrical conditions of non-magnetic atoms found between them.  The order of magnitude of DM interaction, $D$, is therefore, perturbation theory obliges, very small. The explicit form of the DM interaction will be given in the next section. However, in many recent papers using the DM interaction, the assumption of small $D$ is not always respected.  Therefore, we can think that the demonstration of Moriya \cite{Moriya} is a special case and the general Hamiltonian may have the same form but different microscopic origin. This is similar the case of the Hubbard model which was initially originated from a second-order perturbation of exchange interaction, but it has been used with liberty for arbitrary ratio $t/U$.  This is also the case of the Ising model if we think of it as a limiting case of the Heisenberg model.

The DM interaction has been shown to generate skyrmions in various kinds of crystals. For example, it can generate a crystal of skyrmions in which skyrmions arrange themselves in a periodic structure \cite{Rosales, Buhrandt,Kwon1,Kwon2}. Skyrmions have been shown to exist in crystal liquids \cite{Bogdanov2003,Leonov,Ackerman} as well as quantum Hall systems. A single skyrmion has also been found. Effects of skyrmions have been investigated in thin films \cite{Ezawa,Yu2}. Artificial skyrmion lattices have been devised for room temperatures \cite{Gilbert}. Experimental observations of skyrmion lattices have been realized in MnSi in 2009 \cite{Muhlbauer,Bauer} and in doped semiconductors in 2010 \cite{Yu1} .  Needless to say, many potential applications using properties of skyrmions are expected in the years to come.  At this stage, it should be noted that skyrmion crystals can also be created by competing exchange interactions without DM interactions \cite{Hayami,Okubo}. So, mechanisms for creating skyrmions are multiple.

We note that spin-wave excitations in systems with a DM interaction in the helical phase without skyrmions have been investigated by many authors \cite{Diep2017,Puszkarski,Zakeri,Wang,Stashkevich,Moon}.

In this paper, we study a skyrmion crystal created by the competition between the nearest-neighbor (NN) ferromagnetic interaction $J$ and the DM interaction of magnitude $D$ under an applied magnetic field $\vec H$. We show by Monte Carlo (MC) simulation that the skyrmion crystal is stable at finite temperatures up to a transition temperature $T_c$ where the topological structure of each skyrmion and the periodic structure of skyrmions are destroyed.

The paper is organized as follows. Section \ref{Model} is devoted to the description of the model and the method to determine the ground state (GS). It is shown that our model generates a skyrmion crystal with a perfect periodicity at temperature $T=0$. The GS  phase diagram in the space $(D,H)$ is presented. Results showing the stability of the skyrmion crystal at finite $T$ obtained from MC simulations are shown in section \ref{stability}. We show in this section that the relaxation of the skyrmions is very slow and follows a stretched exponential law. The stability of the skyrmion phase is destroyed at a phase transition to the paramagnetic state.  Concluding remarks are given in section \ref{Concl}.

\section{Model and Ground State}\label{Model}
The DM interaction between two spins $\mathbf S_i$ and $\mathbf S_j$ is written as
\begin{equation}\label{eq1}
 \mathbf D_{i,j}\cdot \mathbf S_i\wedge\mathbf S_j
\end{equation}
where $\mathbf D_{i,j}$ is a vector which results from the displacement of non magnetic ions located between $\mathbf S_i$ and $\mathbf S_j$,  for example in Mn-O-Mn bonds in the historical papers \cite{Dzyaloshinskii,Moriya}. The direction of  $\mathbf D_{i,j}$ depends on the symmetry of the displacement \cite{Moriya}.  For two spins, the DM interaction is antisymmetric with respect to the inversion symmetry.

Theoretical and experimental investigations on the effect of the DM interaction in various materials have been extensively carried out in the context of weak ferromagnetism observed in perovskite compounds (see references cited in Refs. \onlinecite{Sergienko,Ederer}). As said in the Introduction, the interest in the DM interaction goes beyond the weak ferromagnetism and the model is used beyond the perturbation limit. It has been shown that the DM interaction is at the origin of topological skyrmions \cite{Muhlbauer,Yu2,Yu1,Maleyev,Lin,Bogdanov,Rossler,Seki,Adams,Heurich,Wessely,Jonietz} and new kinds of magnetic domain walls \cite{Heide,Rohart}.  The increasing interest in skyrmions results from the fact that skyrmions may play an important role in the electronic transport which is at the heart of technological application devices \cite{Fert2013}.

In this paper, we consider for simplicity the two-dimensional case where the spins are on a square lattice in the $xy$ plane. We are interested in the stability of the skyrmion crystal generated in a system of spins interacting with each other via a DM interaction and a symmetric isotropic Heisenberg exchange interaction in an applied field perpendicular to the $xy$ plane. All interactions are limited to NN.
The full Hamiltonian is given by

\begin{eqnarray}
\mathcal{H}&=&-J \sum_{\langle ij \rangle} \vec{S_i} . \vec{S_j} +D \sum_i \overrightarrow{S_i} \wedge (\overrightarrow{S}_{i+x} . \widehat{x} +\overrightarrow{S}_{i+y} . \widehat{y})\nonumber\\
&&-H \sum_i S_i^z
\end{eqnarray}
where the DM interaction and the exchange interaction are taken between NN on both $x$ and $y$ directions.  Rewriting it it in a convenient form, we have
\begin{eqnarray}
\mathcal{H}&=&-J\sum_{\langle ij \rangle} \vec{S_i} . \vec{S_j} +D \sum_i [S_i^y S_{i+x}^z-S_i^z S_{i+x}^y-S_i^x S_{i+y}^z\nonumber\\
&&+S_i^z S_{i+y}^x]- H\sum_i S_i ^z\nonumber\\
&=&-J \sum_{\langle ij \rangle} \vec{S_i} . \vec{S_j} +D \sum_i [S_i^y (S_{i+x}^z-S_{i-x}^z)\nonumber\\
&&-S_i^z (S_{i+x}^y-S_{i-x}^y)-S_i^x (S_{i+y}^z-S_{i-y}^z)\nonumber\\
&&+S_i^z( S_{i+y}^x- S_{i-y}^x)]
- H \sum_i S_i^z
\end{eqnarray}
For the $i$-th spin, one has
\begin{equation}\label{LF}
\mathcal{H}_i= -S_i^x H_i^x - S_i^y H_i^y - S_i^z H_i^zy \nonumber\\
\end{equation}
where the local-field components are given by
\begin{eqnarray}
H_i^x&=&J \sum_{NN} S_j^x+ D(S_{i+y}^z-S_{i-y}^z) \nonumber\\
H_i^y&=&J \sum_{NN} S_j^y- D(S_{i+x}^z-S_{i-x}^z) \nonumber\\
H_i^zz&=&J \sum_{NN} S_j^z+ D(S_{i+x}^y-S_{i-x}^y)- D(S_{i+y}^x-S_{i-y}^x)+H\nonumber
\end{eqnarray}

To determine the ground state (GS), we minimize the energy of each spin, one after another. This can be numerically achieved as the following. At each spin, we calculate its local-field components acting on it from its NN using the above equations. Next we align the spin in its local field, i. e. taking $S_i^x=H_i^x/\sqrt{H_i^x**2+H_i^y**2+H_i^z**2}$ etc. The denominator is the modulus of the local field. In doing so, the spin modulus is normalized to be 1. As seen from Eq. (\ref{LF}), the energy of the spin $\vec S_i$ is minimum. We take another spin and repeat the same procedure until all spins are visited. This achieves one iteration. We have to do a sufficient number of iterations until the system energy converges. For the skyrmion case, it takes about one thousand iterations to have the fifth-digit convergence.

\begin{figure}[h!]
\center
\includegraphics[width=7cm]{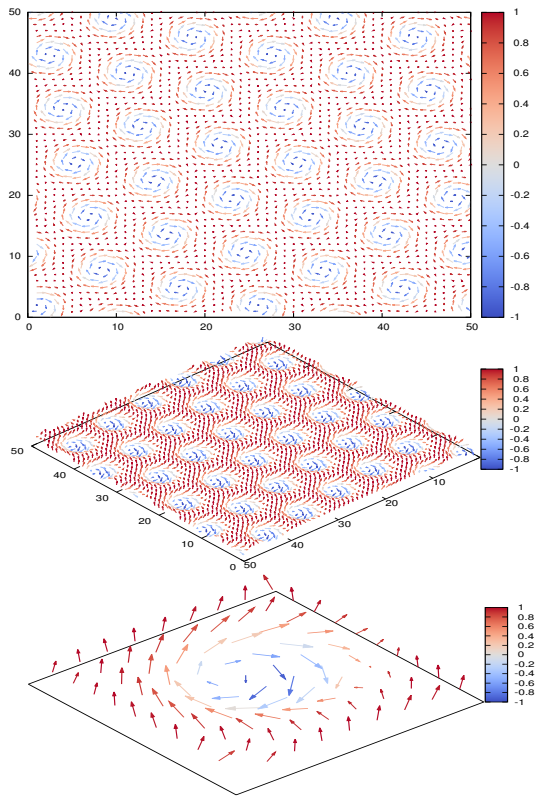}
\caption{Ground state for $D/J=1$ and $H/J=0.5$, a crystal of skyrmions is observed. Top: Skyrmion crystal viewed in the $xy$ plane, Middle: a 3D view, Bottom:  zoom of the structure of a single vortex. The value of $S_z$ is indicated on the color scale. See text for comments. \label{GS1}}
\end{figure}

 An example of  GS are shown in Fig. \ref{GS1}:  a crystal of skyrmions is seen using $D=1$ and $H=0.5$ (in unit of $J=1$).

\begin{figure}[h!]
\center
\includegraphics[width=6cm]{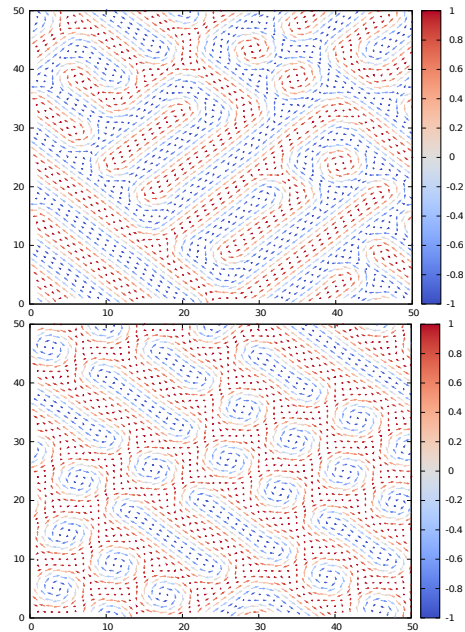}
\caption{Top: Ground state for $D/J=1$ and $H/J=0$, a mixing of domains of long and round islands, Bottom: Ground state for $D/J=1$ and $H/J=0.25$, a mixing of domains of long islands and vortices. We call these structure the "labyrinth phase". \label{GS2}}
\end{figure}

In Fig. \ref{GS2}a we show a GS  at $H=0$ where domains of long and round islands of up spins separated by labyrinths of down spins are mixed.  When $H$ is increased, vortices begin to appear. The GS is a mixing of long islands of up spins and vortices as seen in Fig. \ref{GS2}b obtained with $D=1$ and $H=0.25$.  This phase can be called "labyrinth phase" or "stripe phase".

It is interesting to note that skyrmion crystals with texture similar to those shown in Figs. \ref{GS1} and \ref{GS2} have been experimentally observed in various materials \cite{Yu2,Gilbert,Muhlbauer,Yu3}, but the most similar skyrmion crystal was observed in two-dimensional Fe(0.5)Co(0.5)Si by Yu {\it et al.} using Lorentz transmission electron microscopy \cite{Yu1}.

\begin{figure}[h!]
\center
\includegraphics[width=6cm]{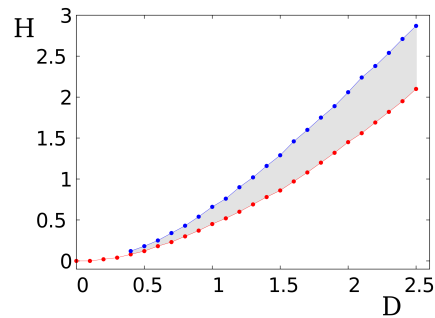}
\caption{Phase diagram in the $(D,H)$ plane for size $N=100$.\label{PD}}
\end{figure}

We have performed the GS calculation taking many values in the plane $(D,H)$.  The phase diagram is established in Fig. \ref{PD}. Above the blue line is the field-induced ferromagnetic phase. Below the red line is the labyrinth phase with a mixing of skyrmions and rectangular domains. The skyrmion crystal phase is found in a narrow region between these two lines.

In the following section, we are interested in the stability of the skyrmion crystal phase as the temperature ($T$) is increased from zero.

\section{Stability of Skyrmion Crystal at Finite Temperatures}\label{stability}

In this section, we show results obtained from MC simulations on a sheet of square lattice of size $N\times N$ with periodic boundary conditions. The first step is to determine the GS spin configuration by minimizing the spin energy by iteration as described above. Using this GS configuration, we heat the system from $T=0$ to a temperature $T$ during an equilibrating time $t_0$ before averaging physical quantities over the next $10^6$ MC steps per spin. The time $t_0$ is the "waiting time" during which the system relaxes before we perform averaging during the next $t_a$.

The definition of an order parameter for a skyrmion crystal is not obvious. Taking advantage of the fact that we know the GS, we define the order parameter as the projection of an actual spin configuration at a given $T$ on its GS and we take the time average.  This order parameter is thus defined as
\begin{equation}\label{OP}
M(T)=\frac{1}{N^2(t_a-t_0)}\sum_i |\sum_{t=t_0}^{t_a} \vec S_i (T,t)\cdot S_i^0(T=0)|
\end{equation}
where $\vec S_i (T,t)$ is the $i$-th spin at the time $t$, at temperature $T$, and $\vec S_i (T=0)$ is its state in the GS. The order parameter $M(T)$ is close to 1 at very low $T$ where each spin is only weakly deviated from its state in the GS. $M(T)$ is zero when every spin strongly fluctuates in the paramagnetic state.
The above definition of $M(T)$ is similar to the Edward-Anderson order parameter used to measure the degree of freezing in spin glasses \cite{Mezard}: we follow each spin with time evolving and take the spatial average at the end.

\begin{figure}[h!]
\center
\includegraphics[width=8cm]{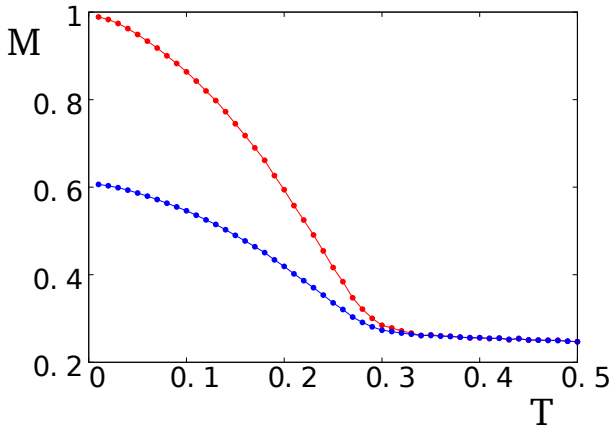}
\caption{Red circles: order parameter defined in Eq. (\ref{OP}) versus $T$, for $H=0.5$ and $N=1800$, averaged during $t_a=10^5$ MC steps per spin after an equilibrating time $t_0=10^5$ MC steps. Blue crosses: the projection
of the $S_z$ on $S_z^0$ of the ground state as defined in Eq. (\ref{OP}) but for the $z$ components only. See text for comments. \label{OPT}}
\end{figure}

We show in Fig. \ref{OPT} the order parameter $M$ versus $T$ (red data points) as well as the average $z$ spin component (blue data points) calculated by the projection procedure for the total time $t=10^5+10^6$ MC steps per spin.  As seen, both two curves indicate a phase transition at $T_c\simeq 0.26 J/k_B$. The fact that $M$ does not vanish above $T_c$ is due to the effect of the applied field. It should be said that each skyrmion has a center with spins of negative $z$ components (the most negative at the center), the spins turn progressively to positive $z$ components while going away from the center.

\begin{figure}[h!]
\center
\includegraphics[width=6cm]{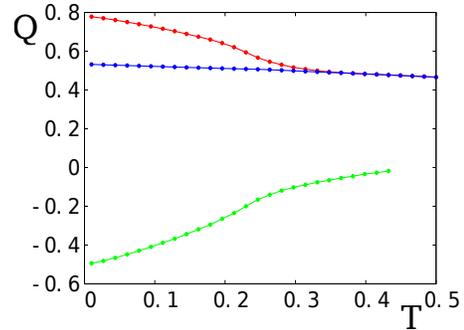}
\caption{Order parameters defined in Eqs. (\ref{EA1})-(\ref{EA2}) versus $T$, for $H=0.5$ and $N=800$, $t_0=10^5$, $t_a=10^6$.  Red circles: total positive $z$ component ($Q_+(T)$). Green circles: total negative $z$ component $Q_-(T)$.  Blue circles: total $z$ component. See text for comments.\label{EAT}}
\end{figure}

We can also define another order parameter: since the field acts on the $z$ direction,  in the GS and in the skyrmion crystalline phase we have both positive and negative $S_z$. In the paramagnetic state, the negative $S_z$ will turn to the field direction.  We define thus the following parameters using the $z$ spin-components
\begin{eqnarray}
Q_+(T)&=&\frac{1}{N^2(t_a-t_0)}\sum_{S_i^z>0} \sum_{t_0}^{t_a}  S_i^z (T,t)\label{EA1}\\
Q_-(T)&=&\frac{1}{N^2(t_a-t_0)}\sum_{S_i^z<0} \sum_{t_0}^{t_a}  S_i^z (T,t)\label{EA2}
\end{eqnarray}
Figure \ref{EAT} shows $Q_+$ and $Q_-$ versus $T$. As seen, at the transition $Q_+$ undergoes a change of curvature and $Q_-$ becomes zero. All spins have positive $S_z$ after the transition due to spin reversal by the field.

The results obtained at the end of the simulation may depend on the overall time $t=t_0+t_a$.  In simple systems, the choices of $t_0$ and $t_a$ can be guided by testing the time-dependence of physical quantities, and the values of $t_0$ and $t_a$ are chosen when physical quantities do not depend on these run times. However, in disordered systems such as spin glasses and in complicated systems such as frustrated systems, the relaxation time is very long and out of the reach of simulation time. In such cases, we have to recourse to some scaling relations in order to deduce the values of physical quantities at equilibrium \cite{Phillips,Ogielski}.  We show below how to obtain the value of an order parameter at the infinite time.

\begin{figure}[h!]
\center
\includegraphics[width=8cm]{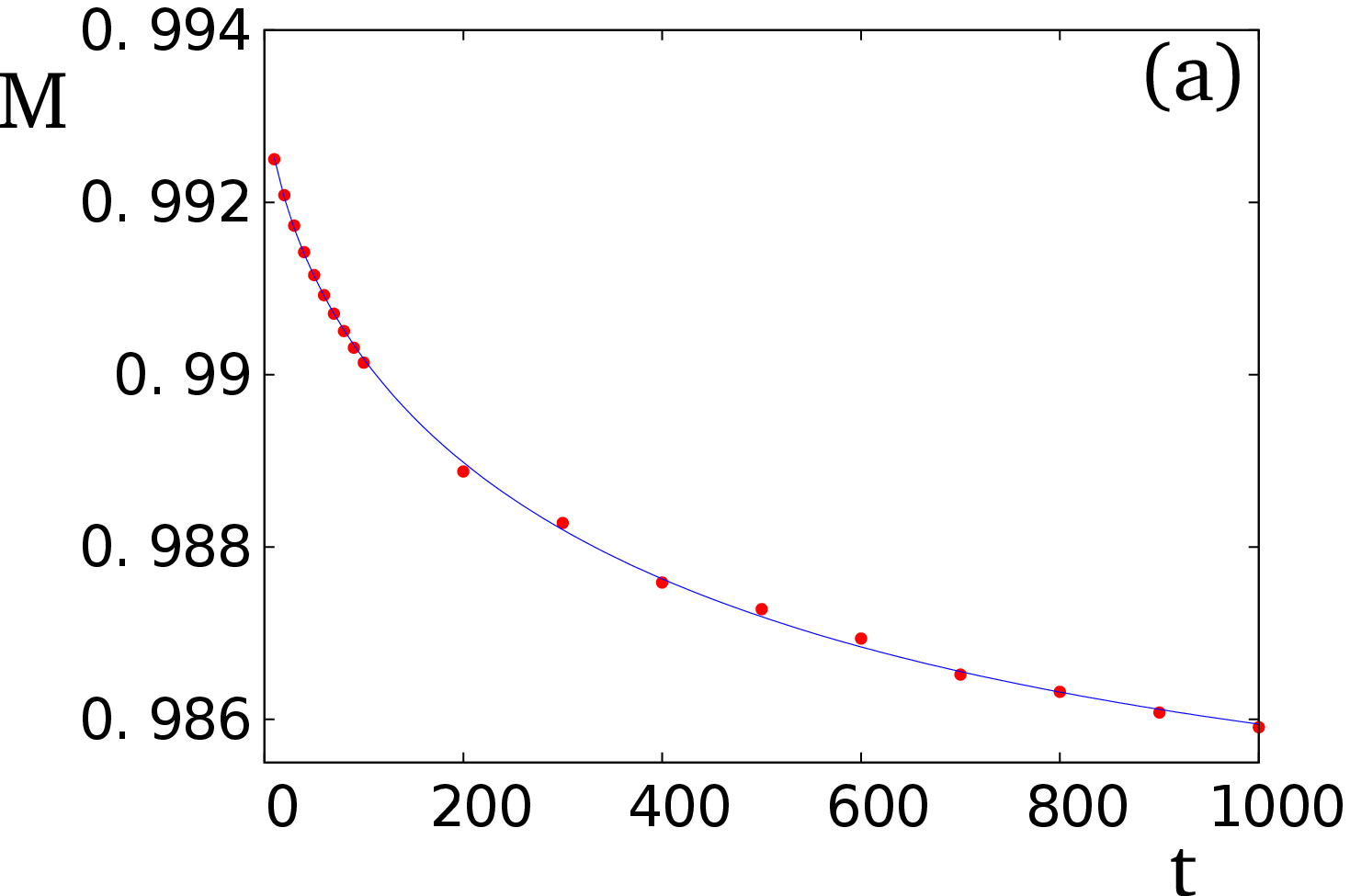}
\includegraphics[width=8cm]{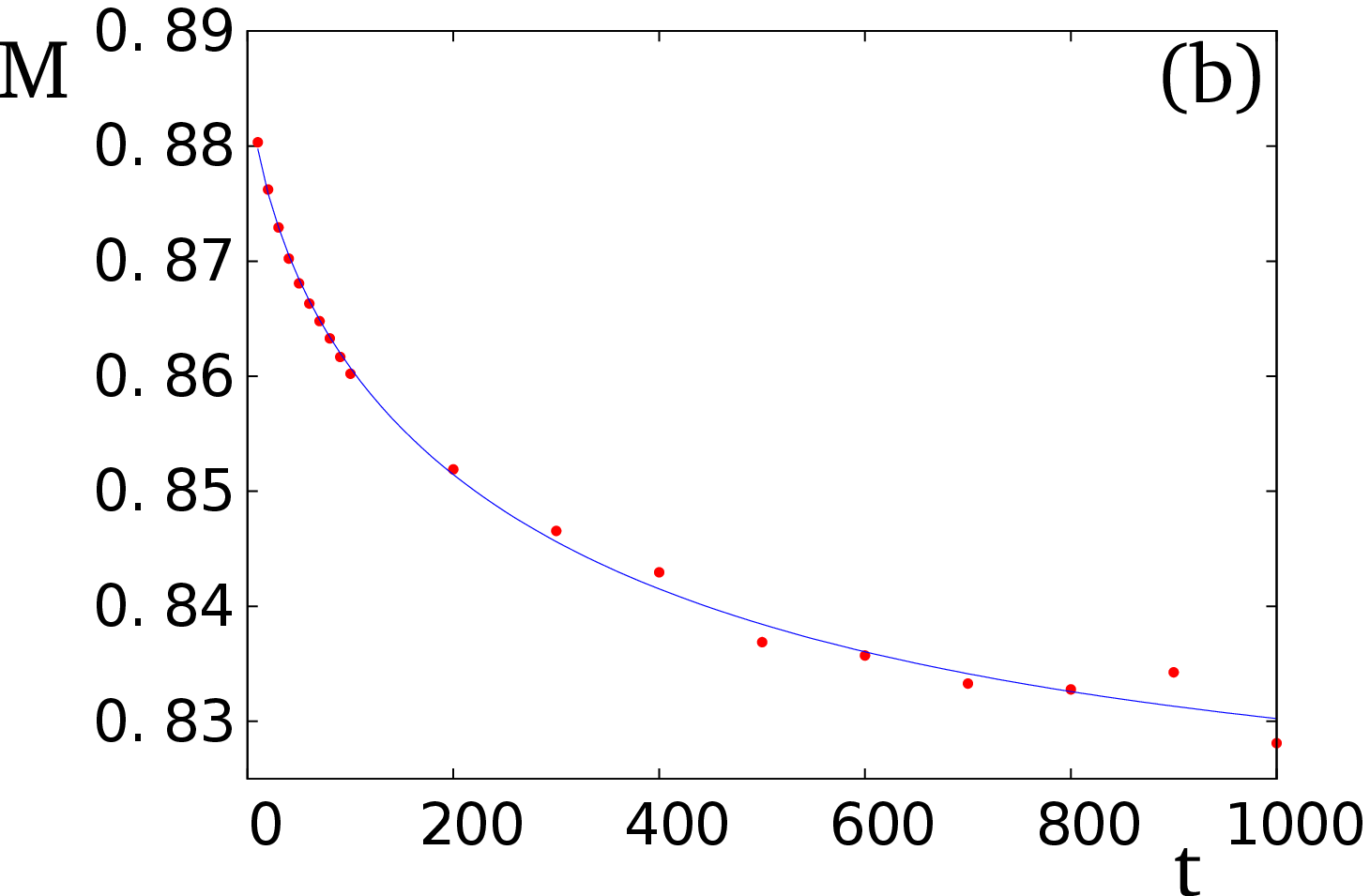}
\includegraphics[width=8cm]{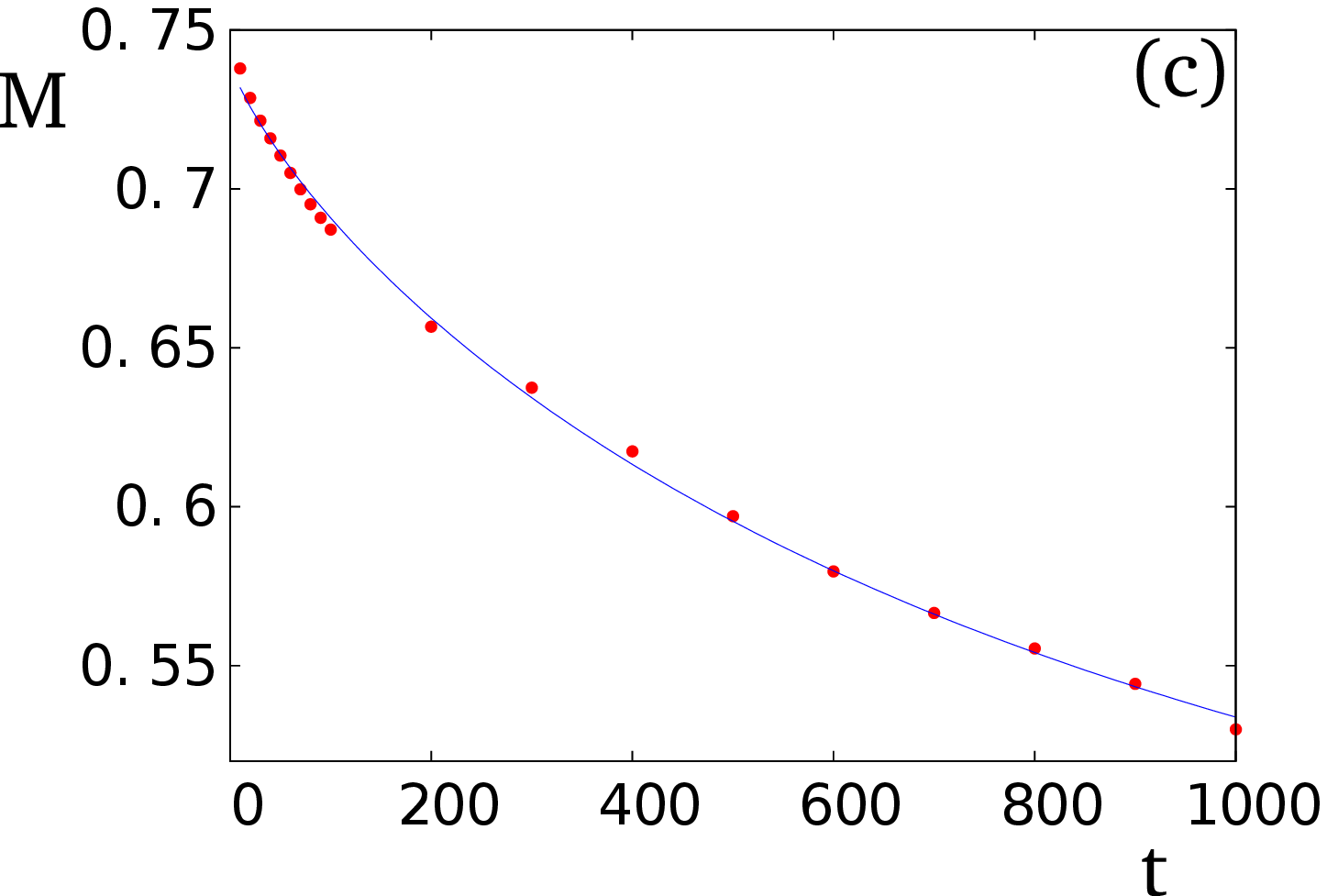}
\caption{The order parameter $M$ defined by Eq. (\ref{OP})  versus MC time $t$ in unit of 1000 MC steps per spin, for $H=0.5$ and $N=800$ (a) $T=0.01$, values of fitting parameters $\alpha=0.6$, $A=0.008\pm 0.00001$, $\tau=(364\pm 19)10^3$, $c=0.984505\pm0.00012$; (b) $T=0.094$, $\alpha=0.6$, $A=0.0653 \pm 0.0013$, $\tau =(277 \pm 36)10^3$, $c=0.822 \pm 0.018)$; (c) $T=0.17$, $\alpha=0.8$, $A=0.31\pm 0.01$, $\tau=891 \pm 100)10^3$, $c=0.43\pm0.017$.\label{relax}}
\end{figure}

In order to detect the dependence of $M(T)$ on the total MC time $t=t_0+t_a$, we calculate the average of $M(T)$ over $10^6$ MC steps per spin, after a waiting time $t_0$ as said above. We record the values of $M(T)$ in different runs with $t_0$ varying from $10^4$ to $10^6$ MC steps per spin.  We plot these results as a function of different total time $t$ in Fig. \ref{relax} for three temperatures.  As said above, to find the value extrapolated at the infinite time,
we use the stretched exponential relaxation defined by
\begin{equation}
\label{eq:srf}
M(T,t)=A\exp\left[-(t/\tau)^{\alpha}\right]+c,
\end{equation}
where $t$ is the total simulation time,  $\alpha$ is the stretched exponent, $A$ a temperature-dependent constant, and $\tau$  the  relaxation time.  Note that this definition, without the constant $c$, has been used by many previous authors in the context of spin glasses \cite{DeDominicis,Almeida,Campbell2011,Suzuki,Malinowski2011}. We have introduced $c$ which is the infinite-time limit of $M(T)$. We have taken  $t$ from $10^4$ to $10^6$ MC steps per spin in the simulation. At the infinite-time limit, $c$ is zero for $T\gg T_c$, and $c\neq 0$ for $T<T_c$.  Figure \ref{relax} shows $M(T,t)$ as a function of time $t$ in unit of $10^3$ MC steps per spin, for three temperatures $T=0.01$, 0.094 and 0.17.  As seen, the fit with Eq. (\ref{eq:srf}) presented by the continuous line is very good for the whole range of $t$.

Several remarks are in order:

(i) the precision of all parameters are between 1\% to 5\% depending on the parameter,

(ii) the value of $\alpha$ can vary a little bit according to the choice and the precision of the other parameters in the fitting but this variation is within a very small window of values around the value given above. For example, at $T=0.17$, $\alpha$ can only be in the interval $[0.8\pm 0.02]$. The value of $\alpha$ can vary with temperature as seen here: at low $T$, $\alpha=0.6$, and at a higher $T$, we have $\alpha=0.8$. This variation has been seen in other systems, in particular in spin glasses \cite{Ngo2014}.

(iii)  the relaxation time, within statistical errors, is approximatively constant at low $T$, but it increases rapidly when $T$ tends to $T_c$ as seen in the value of $\tau$ at $T=0.17$. This increase is a consequence of the so-called "critical slowing-down" when the system enters the critical region.

\begin{figure}[h!]
\center
\includegraphics[width=8cm]{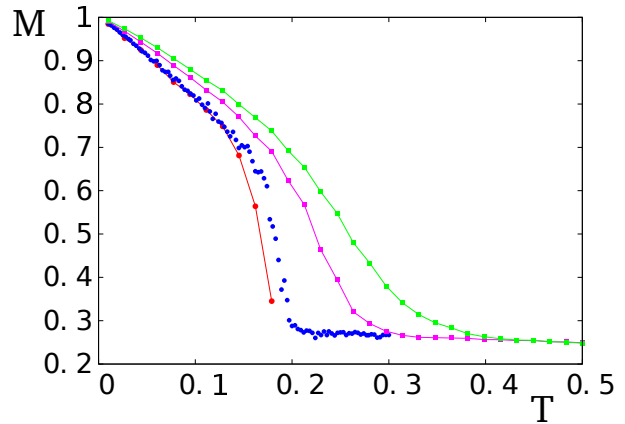}
\caption{The order parameter $M(T)$  versus $T$ for several waiting times $t$, for $H=0.5$ and $N=800$: from above $t=10^5$, $2\times10^5$, $10^6$, $\infty$ (by fitting with Eq. (\ref{eq:srf}).\label{MT}}
\end{figure}

Let us show $M(T)$ as a function of $T$ in Fig. \ref{MT} using the results of different run times from $t=10^4+10^6$ MC steps per spin to $t=10^6+10^6$.  The values of at infinite time for each $T$ deduced from Eq. (\ref{eq:srf}) is also shown.
We see that while the total time $10^5+10^6$ MC steps per spin is sufficient at low $T$, it is not enough at higher $T$. That was the reason why we should use Eq. (\ref{eq:srf}) to find the value of $M(T)$ at infinite time to be sure that the skyrmion crystal is stable at finite temperatures.

We have studied finite-size effects on the phase transition at $T_c$ and we have seen that from $N=800$, all curves coincide: there is no observable finite size effects for $N\geq 800$. The present 2D skyrmion lattice phase therefore remains stable for the infinite dimension, unlike the ferromagnetic Heisenberg model in 2D \cite{Mermin}. Note that the nature of the ordering at low $T$ and the phase transition observed here, in spite of the vortex nature of skyrmions,  are not those of the Kosterlitz-Thouless vortex mechanism observed in ferromagnetic XY spin systems \cite{KT} since our "vortices" are stable without changing their position up to the transition temperature.

\section{Concluding Remarks}\label{Concl}
In this paper, we have shown that the competition between a ferromagnetic interaction $J$ and a Dzyaloshinskii-Moriya interaction $D$ under an applied magnetic field $H$ in two dimensions generate a skyrmion crystal in a region of the phase space $(D,H)$.  The spin model is the classical Heisenberg model. We have numerically determined the ground state by minimizing the local energy, spin by spin, using an iteration procedure.  The skyrmion lattice is then heated to a finite temperature by the use of Monte Carlo simulations. We have shown that the skyrmion lattice is stable up to a finite temperature $T_c$ beyond which the system becomes disordered.    We have also shown that the relaxation follows a stretched exponential law. We believe that such a stability can be experimentally observed in real systems.

Perhaps, our model is the simplest model able to generate a skyrmion crystal without the need to include more complicated interactions such as long-range dipolar interactions, easy and uniaxial anisotropies \cite{Kwon2,Lin,Yu3}.  Note that other authors have shown that skyrmion crystals can also be generated only with frustrated short-range interactions without even the DM interaction \cite{Hayami,Okubo}.  So, we have to keep in mind that there are many interaction mechanisms of different nature which can generate skyrmion crystals. The origin of a skyrmion crystal determines its structure, its stability, its dynamics and in short its properties (see a discussion on this point in \cite{Yu3}).  Experiments can be used therefore to determine the interaction mechanism which is at the origin of the formation of a skyrmion crystal.

\acknowledgments
SEH acknowledges  a financial support from Agence Universitaire de la Francophonie (AUF).

{}

\end{document}